\documentclass[12pt,fleqn]{article}
\usepackage{amsmath,amssymb}
\usepackage{bm}
\usepackage[dvips]{graphicx}
\allowdisplaybreaks[3]

\newcommand{\be}{\begin{equation}}
\newcommand{\ee}{\end{equation}}
\newcommand{\ba}{\begin{eqnarray}}
\newcommand{\ea}{\end{eqnarray}}

\begin{document}

\title{On the nature of the $\Lambda(1405)$ as a superposition of two states.}

\author{E. Oset$^a$, V.K. Magas$^b$ and Ramos$^b$}
  
\maketitle 

\begin{center}
\textit{$^a$ Departmento de F\'isica Te\'orica and IFIC,
Centro Mixto Universidad de Valencia-CSIC,
Institutos de Investigaci\'on de Paterna, Aptd. 22085, 46071
Valencia, Spain}\\

\textit{$^b$ Departament d'Estructura i Constituents de la Mat\'eria,
Universitat de Barcelona,
 Diagonal 647, 08028 Barcelona, Spain} \\

\end{center}

\vspace{0.4cm}

\abstract{
 We use recent data on the $K^- p \to \pi^0 \pi^0 \Sigma^0$ reaction with the 
$\pi^0 \Sigma^0$ mass distribution of 
forming the $\Lambda(1405)$ with a peak at $1420$ MeV and a relatively
narrow width of $\Gamma = 38$ MeV, together with  
 those of the $\pi^- p \to K^0 \pi
\Sigma$ reaction to show that there are
two $\Lambda(1405)$ states instead of one as so far assumed. }

\vspace{2cm}

The  $\Lambda(1405)$ has been described as a dynamical resonance 
generated
from the interaction of meson baryon components in coupled channels by means of 
unitary extensions of chiral perturbation theory ($U\chi PT$) 
\cite{Kaiser:1995cy,kaon,Oller:2000fj,Jido:2002yz,Garcia-Recio:2002td}.
The surprise, however, came with the
realization that there are two poles in the neighborhood of the 
$\Lambda(1405)$ both contributing to the final experimental invariant mass
distribution \cite{Oller:2000fj,Jido:2002yz,Garcia-Recio:2002td,Jido:2003cb,
Garcia-Recio:2003ks,Hyodo:2002pk,Nam:2003ch}.
The properties of these two states are quite different, one has a mass around 
$1390$ MeV, a large
width of about $130$ MeV and couples mostly to $\pi \Sigma$, while the second
one has a mass around $1425$ MeV, a narrow width of about $30$ MeV and couples
mostly to $\bar{K} N$. The  two states are populated
with different weights in different reactions and, hence, their superposition can lead to
different distribution shapes. Since the $\Lambda(1405)$ resonance is always seen from the
invariant mass of its only strong decay channel, the $\pi \Sigma$,
hopes to see the second pole are tied to having a reaction where
the $\Lambda(1405)$ is formed from the $\bar{K} N$ channel. This is accomplished
by the recently measured reaction
$K^- p \to \pi^0 \pi^0 \Sigma^0$  \cite{Prakhov} which allows us to test already the
two-pole nature of the $\Lambda(1405)$.


\begin{figure}[htb]
\begin{center}
   \includegraphics[height=3.0cm,width = 5.5cm]{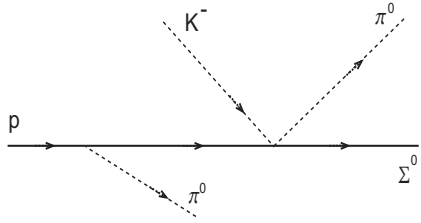}
 \caption{\label{fig:tree}
Nucleon pole term for the $K^- p \to \pi^0 \pi^0 \Sigma$ reaction.}
\end{center}
\end{figure}


Our model for the reaction 
$K^- p \to \pi^0 \pi^0 \Sigma^0 $
in the energy region of $p_{K^-}=514$ to $750$ MeV/c, as in the experiment \cite{Prakhov}, 
considers those mechanisms in which a $\pi^0$ loses the necessary energy
to allow the remaining $\pi^0\Sigma^0$ pair to be on top of the $\Lambda(1405)$
resonance.  The first of such mechanisms is given by the diagram of
Fig.~\ref{fig:tree}. Other mechanisms that involve the meson meson interaction
and baryon-baryon-three meson vertices were found negligible in the detailed 
study of \cite{prl}.

\begin{figure*}[htb]
\begin{center}
  \includegraphics[width = 16cm]{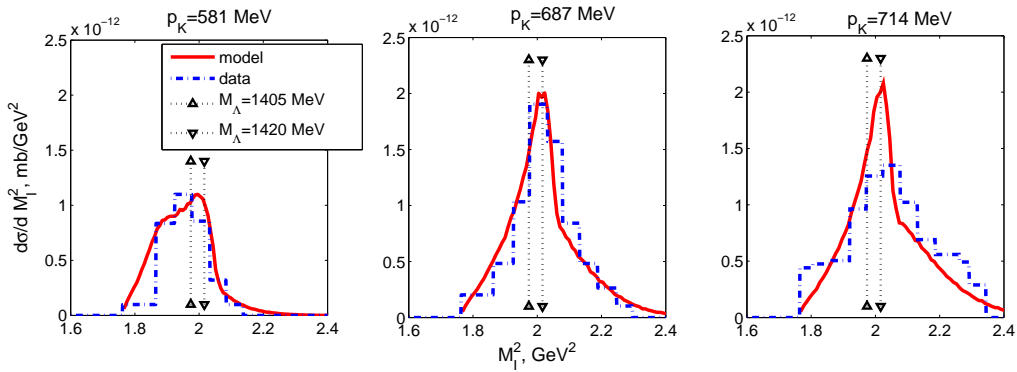}
 \caption{The ($\pi^0 \Sigma^0$) invariant mass distribution for three different initial kaon
momenta.
 \label{fig:mass}
 }
\end{center}
\end{figure*}


The indistinguishability of the two emitted pions requires the implementation
of symmetrization. This is achieved by summing two amplitudes evaluated
with the two pion momenta exchanged. In addition,
a factor of 1/2 for indistinguishable particles is also included in the
total cross section.

Our calculations show that the process is largely dominated by the nucleon
pole term shown in Fig.~\ref{fig:tree}. As a consequence, the $\Lambda(1405)$ thus obtained comes
mainly from the $K^- p \to \pi^0 \Sigma^0$ amplitude which, as mentioned above,
gives  the largest possible weight  to the second (narrower) state.


\begin{figure}[htb]
\begin{center}
  \includegraphics[width = 8.0cm]{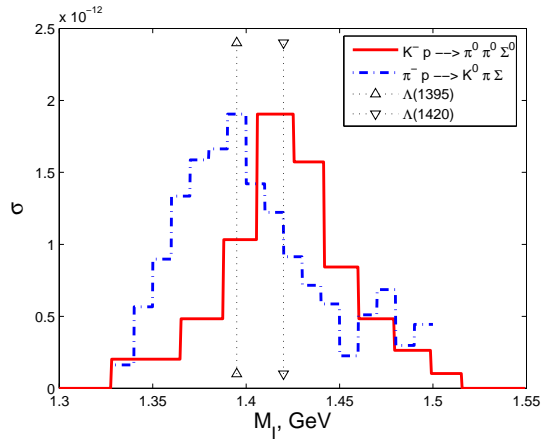}
 \caption{Two experimental shapes of  $\Lambda(1405)$ resonance. 
 See text for more details. 
 \label{two_exp}
 }
\end{center}
\end{figure}


In Fig.~\ref{fig:mass} our results for the invariant mass distribution
for three different energies of the incoming $K^-$ are
compared to the experimental data. Symmetrization of the amplitudes produces a
sizable amount of background. At a kaon laboratory momentum of $p_K=581$ MeV/c
this background  distorts the $\Lambda(1405)$ shape producing cross section in
the lower part of $M_I$, while at $p_K=714$ MeV/c the strength of this
background is shifted toward the higher $M_I$ region. An ideal situation is
found for momenta around $687$ MeV/c, where the background sits below the
$\Lambda(1405)$ peak distorting its shape minimally. The peak of the resonance
shows up at $M_I^2=2.02$ GeV$^2$ which corresponds to $M_I=1420$ MeV, larger
than the nominal $\Lambda(1405)$, and in agreement with the predictions of
Ref.~\cite{Jido:2003cb} for the location of the peak when the process is
dominated by the $t_{{\bar K}N \to \pi\Sigma}$ amplitude.  The apparent width
from experiment is about $40-45$ MeV, but a theoretical analysis
permits extracting the pure resonant part by not symmetrizing the amplitude 
\cite{prl} and one finds $\Gamma=38$ MeV. This is
 smaller than the nominal $\Lambda(1405)$ width of $50\pm 2$ MeV \cite{PDG},
obtained from the average of several experiments, and much narrower than the
apparent width of about $60$ MeV that one sees in the $\pi^- p \to K^0 \pi
\Sigma$ experiment \cite{Thomas}, which also produces a distribution peaked at
$1395$ MeV.
In order to illustrate the difference between the $\Lambda(1405)$ resonance
seen in this latter reaction and in the present one, the two
experimental distributions are compared in Fig. \ref{two_exp}. We recall
that the shape of the  $\Lambda(1405)$ in the $\pi^- p \to K^0 \pi \Sigma$ 
reaction was shown in Ref.~\cite {hyodo} to be largely built from the
 $\pi \Sigma \to \pi \Sigma$ amplitude, which is dominated by
the wider lower energy state.

The invariant mass distributions shown here are not normalized, as in
experiment. But we can also compare our absolute values of
the total cross sections with those in Ref.~\cite{Prakhov}. This is done in Ref.
\cite{prl} and the agreement is good.

 The study of the present
 reaction, complemental to the one of Ref.  \cite {hyodo} for the $\pi^- p \to
 K^0 \pi \Sigma$ reaction, has shown that the quite different shapes of the 
 $\Lambda(1405)$ resonance seen in these experiments can be interpreted in favour
 of the existence of two poles with the
 characteristics predicted by the chiral theoretical calculations.  
 Besides demonstrating once more the great predictive power of the chiral
 unitary theories, this combined study of the two reactions gives the first
 clear evidence of the two-pole nature of the $\Lambda(1405)$. 

 \vspace*{1cm}

{\bf Acknowledgments:}\ \ 
This work is partly supported by DGICYT contracts BFM2002-01868, BFM2003-00856,
the Generalitat de Catalunya contract SGR2001-64,
and the E.U. EURIDICE network contract HPRN-CT-2002-00311.
This research is part of the EU Integrated Infrastructure Initiative
Hadron Physics Project under contract number RII3-CT-2004-506078.


\end{document}